\documentclass[final]{aipproc}
\layoutstyle{6x9}
\usepackage{graphicx}
\usepackage{natbib}

\begin{document}

\title{Cosmic Magnetic Fields: from Stars and Galaxies to the Primordial Universe}

\classification{96.12.Hg; 97.10.Gz; 97.10.Ld; 98.35.Eg; 98.38.Dq; 98.38.Fs; 98.52.-b; 98.54.Cm; 98.58.Kh; 98.58.Mj; 98.62.En; 98.65.-r; 98.65.Hb; 98.80.Bp}

\keywords{Magnetic fields in: stars,  interstellar medium, jets and accretion disks, galaxies, clusters of galaxies,  intracluster and inter-galactic medium, and in the early universe.}

\author{Elisabete M. de Gouveia Dal Pino}{
  address={Instituto Astronômico e Geofísico - Universidade de São Paulo (IAG-USP); dalpino@astro.iag.usp.br
}
}

\begin{abstract}
Most of the baryonic matter in the Universe is permeated by magnetic fields which affect many, if not most, of astrophysical phenomena both, in compact sources and in diffuse gas. Recent years have been marked by a worldwide surge of interest in the astrophysical magnetic fields, their origin, and their influence on the formation and evolution of astrophysical objects (stars, galaxies, cooling flows). This growing interest is in part due to the fact that it has become possible to trace magnetic fields in molecular clouds, over vast extensions of the Milky Way and to study extragalactic magnetic fields, including fields in clusters of galaxies. With the combination of various techniques, such as Zeeman and Faraday rotation measurements with  synchrotron and aligned grain polarimetry, it is now possible to undertake quantitative observational studies of magnetic fields, the results of which can be compared with high resolution dynamo and MHD turbulence simulations. This brings the field to a new stage and, at the same time, calls for addressing fundamental questions, such as the correspondence of basic processes in astrophysical media with computer simulations, and our real understanding of the processes that we rely to infer magnetic fields from observations. Presently, there seems to be now no doubts that magnetic fields play a crucial role in star formation, solar and stellar activity, pulsars, accretion disks, formation and stability of jets, and formation and propagation of cosmic rays. They are also probably crucial in the interstellar medium dynamical evolution, supernova remnants, gamma-ray bursts, and the galaxy structure, but its importance is still not well understood in stellar and galaxy evolution, and structure formation in the early Universe where its origin is virtually a mystery. In this paper, I will briefly review the importance of the cosmic magnetic fields both from a theoretical and from an observational perspective, focusing on their role in stellar and compact objects, in the interstellar medium and star formation regions, and in galaxies, clusters of galaxies, and the primordial Universe.
\end{abstract}

\maketitle


\section{INTRODUCTION}

  About half a century ago cosmic magnetic fields were generally regarded as unimportant. Only a few scientists such as Alfvén, Biermann, Chandrasekhar, and Parker, realized the potential role of the magnetism in the Universe. In Brazil, pioneering research on cosmic magnetism started in the Brazilian Center of Theoretical Physics (CBPF) in Rio de Janeiro and in the Institute of Astronomy, Geophysics and Meteorological Sciences of the University of S\~ao Paulo (IAG-USP) only in the sixties and seventies of the last century. In a Lecture in 1987, H.C.  van der Hulst expressed the general uncomfortable  feeling that then permeated (and in some circumstances still does) the Astrophysical community with regard to cosmic magnetic fields: ''...Magnetic Fields are to Astrophysics as sex is to Psychology...''. Since then, our view has changed considerably. We know now that most of the visible matter in the Universe is in a plasma state, or more specifically is composed of ionized or partially ionized gas permeated by magnetic fields. There seems to be now no doubts that magnetic fields play a crucial role in star formation, solar and stellar activity, pulsars, formation and stability of jets and accretion disks, formation and propagation of cosmic rays, and stability of galactic disks. They are also probably crucial in the interstellar medium (ISM) dynamical evolution, in molecular clouds, supernova remnants, proto-planetary disks, and planetary nebulae, but its importance is still not well understood in stellar evolution, halos of galaxies, galaxy evolution, and structure formation in the early Universe (e.g., \cite{Beck2005, de Gouveia Dal Pino2006a}). In this lecture, I will very briefly review the importance of the cosmic magnetic fields, particularly focusing on compact objects, the ISM and star formation regions, and on galaxies, clusters, and the early Universe. I will also briefly discuss the potential origin of the cosmic magnetic fields.

Before starting this review, let me consider the magnetic induction equation that describes the magnetic field evolution in a fluid:

\begin{equation}
\frac{\partial \mathbf{B}}{\partial t} = \nabla \times \left( \mathbf{u} \times \mathbf{B} \right) + \eta_{Ohm} \nabla ^{2} \mathbf{B}
\end{equation}

Where $\eta_{Ohm}$  is the Ohmic resistivity. The first term of the RHS accounts for the flux freezing between the ionized gas and the magnetic field lines and the second one accounts for the magnetic diffusion. The comparison of the second term with the one of the LHS gives the diffusion time scale for the magnetic field configuration decay, $t_{diff}(L)= L^2/\eta_{Ohm}$, where $L$ is the characteristic length scale of the spatial variation of $B$. In most astrophysical systems and environments, the gas is a good conductor and since the typical length scales L are very large, one finds that $t_{diff} (L)$ is generally much longer than the typical dynamical time scales of the systems. This implies that the first term on the RHS (the frozen-in term) generally dominates and the magnetic flux is conserved in most astrophysical plasmas, with the possible exception of magnetic boundary layers or surface discontinuities, like magnetic reconnection regions, where the condition above may be violated. Therefore, as a golden rule in Astrophysics, one may say that once magnetic fields are created in cosmic environments it is very difficult to get rid of them!

\section{HOW DO WE MEASURE COSMIC MAGNETIC FIELDS?}

Cosmic magnetic fields are observable through the optical polarization of (unpolarized) star light by interstellar dust grains that are aligned with the embedded magnetic field in the ISM; through the Zeeman splitting of spectral emission lines from sources by the ambient field; through linearly polarized radio synchrotron emission produced by relativistic particles accelerated in the magnetic fields of some astrophysical sources like pulsars, supernova remnants or radio galaxies; and through the Faraday rotation of the angle between the polarization vector and the line of sight by the ambient magnetic field where the polarized light travels. For more detailed reviews on these observational methods and techniques I refer to Beck \cite{Beck2005,Beck2009},  Wielebinski et al. \cite{Wielebinski et al2009} and references therein.

\section{MAGNETIC FIELDS IN THE SUN AND OTHER STARS}

In the case of the Sun, our nearest and, therefore, most investigated astrophysical source, it is well known that its surface phenomena, such as flares, sunspots and coronal mass ejections (CMEs), are all the result of intense magnetic activity.  Magnetic arcs that extend up to $10^4$ km above the surface arise by buoyancy due to convective motions. The sunspots that are located at the feet of these magnetic arcs have magnetic field intensities that can be as large as 2000 G. Solar flares, which correspond to sudden release of energy of the order of $10^{30}-10^{32}$ erg in a period of seconds to hours, are caused by violent magnetic reconnection of the lines in the magnetic loops (e.g., \cite{Shibata2005}) and the released energy is believed to be responsible for the solar coronal gas heating (up to temperatures of the order of $2 \times 10^6$ K), particle acceleration, and the production of the CMEs although the variety of processes through which these phenomena occur are not yet fully understood. Despite its chaotic appearance, the solar magnetic activity is cyclic, as discovered more than 150 years ago, when  Schwabe (in 1844) detected the periodic time variability in the number of sunspots. The sunspots usually appear in pairs (with opposite polarity) in a belt of latitudes within $30^o$ (in both sides of the solar equator). While the strength of the magnetic fields in the sunspots is around $10^3$ G, the magnitude of the diffuse poloidal field is of tens of G. Its polarity changes every 11 years and the total period of the cycle is 22 years. When the number of sunspots is maximum, the toroidal fields attain its maximum and the polar field inverts its polarity, so that there is a phase lag of $1/2$ between both the toroidal and poloidal components of the magnetic field.

It is generally believed that the solar cycle corresponds to a
hydromagnetic dynamo process operating in some place within the
solar interior.
Parker (1955) was the first to build a solar dynamo model, since then there
has been important developments in the observations, theory and simulations,
but a definitive model for the solar dynamo is still missing. Helioseismology has
mapped the solar internal rotation showing a detailed profile of the latitudinal
and radial shear layers that supports the idea that the first part of the dynamo
process corresponds to the transformation of  poloidal field into
toroidal field through differential rotation. This stage is known as the
 $\Omega$ effect. Usually, it is believed that
this process happens at the base of the convection zone where a strong shear
layer (the tachocline) was found. To close the dynamo cycle, a
 second stage is necessary, i.e., the transformation of the toroidal field
into a new poloidal field of opposite polarity. This is a less understood process. Two main hypotheses have
been formulated in order to explain this phase, usually called the
$\alpha$ effect: the first one is based on the Parker's  idea of a turbulent mechanism
in which the poloidal field is the result of cyclonic motions operating in  small
scales in the toroidal field that would cause the arise of small scale magnetic loops to the surface. These small loops should reconnect to form a large scale dipolar field. The second one is based on the ideas of  Babcock \cite{bab61}
and Leighton \cite{Leighton1969} (BL). They proposed that a tilted bipolar magnetic region
(BMR) has a net dipolar moment, with the leading part moving towards the
equator and its companion going to the pole forming a poloidal structure. An
effective flux-transport mechanism is then necessary to transport the dipolar
regions towards the pole and from there to the internal layers in order to allow
the dynamo to begin again, and so on. Most of the BL models include meridional
flow.
Several numerical studies have been performed in order to simulate this process that have been relatively successful in reproducing the main features of the solar cycle described above (see, e.g. \cite{Dikpati and Charbonneau1999, Guerrero and de Gouveia Dal Pino2007}), however they face difficulties that have been object of criticism in the literature (e.g., \cite{Brandenburg2005,Brandenburg and Subramanian2005}). For instance, magnetic flux tubes formed at the base of the convection zone need to be of the order of $10^5$ G in order to produce the observed BMRs, but this implies an energy density that is one to two orders of magnitude larger than the equipartition value with kinetic energy inthat place. Besides, BL models operate in the flux-transport regime. This means that the advection time dominates upon the diffusion time. This requires that BL models use diffusivity values lower than $10^{11}$ cm/s in the convection zone, but values larger than $10^{12}$ cm/s are observed at the surface. Helioseismic observations inferred a second layer of intense radial shear located just beneath the solar surface \cite{Corbard and Thompson2002} and, as discussed by Brandenburg \cite{Brandenburg2005}, there are reasons that seem to favor a dynamo operating there. Nevertheless, it does not rule out the possibility that the solar dynamo is of BL type; however the presence of the near-surface shear layer may change the way it operates.

In order to solve these difficulties, several solutions have been proposed based on  the addition of  turbulent components to the large scale components of the flow. Guerrero \& de Gouveia Dal Pino \cite{Guerrero and de Gouveia Dal Pino2008} considered the turbulent pumping as an alternative mechanism to advect the magnetic flux, and Guerrero, Dikpati \& de Gouveia Dal Pino \cite{Guerrero et al2009} included the non-linear back-reaction of the magnetic field on the turbulent magnetic diffusivity, a process known as $\eta$-quenching. Guerrero \& de Gouveia Dal Pino \citet{Guerrero and de Gouveia Dal Pino2008} found that it is possible to build a flux-transport dynamo model able to reproduce the observations as long as a thin tachocline located below the convective zone is considered. This helps to prevent the amplification of undesirable strong toroidal fields at the high latitudes. They have have also found that it is important to consider the turbulent magnetic pumping mechanism because it provides magnetic field advection both equatorward and inwards that results in a correct latitudinal and temporal distribution of the toroidal field and also allows the penetration of the toroidal fields down into the stable layers where they can acquire further amplification. Besides, this mechanism plays an important role in reproducing the correct field parity (anti-symmetric) on both solar hemispheres. Guerrero, Dikpati \& de Gouveia Dal Pino \citet{Guerrero et al2009} also found that the $\eta$-quenching may lead to the formation of long-lived small structures of toroidal field which resemble the flux-tubes that are believed to exist at the base of the convection zone. These magnetic fields can be up to  twice as large as the magnetic structures which are developed without this effect.

Finally, a number of theoretical works in the last years have called the attention to the role of magnetic helicity conservation in the dynamo processes, giving new life to the turbulent dynamo model as proposed by Parker. With the aim to study the role of magnetic helicity and explore a more realistic dynamical description of the dynamo mechanism, several authors have  recently developed 3D convective numerical models to try to reproduce the natural scenario of the solar interior where the dynamo might operate (e.g., ). These are however still in their childhood \cite{Brun et al2004, Miesch et al2008, Brandenburg and Subramanian2005, Ossendrijver et al2002, Guerrero and de Gouveia Dal Pino2010}.

Figure 1 shows examples of 3D magneto-convection numerical simulations  performed to reproduce the dynamo action and the alpha-effect arising from convective turbulence.  Two layers were considered originally in hydrostatic equilibrium, one subadiabatic layer at the bottom and one superadiabatic convective layer at the top. The authors have allowed this hydrodynamic model to evolve up to a  steady state and then  introduced a seed magnetic field. These preliminary results indicate that the presence of rotation influences the development of a large scale magnetic field with exponential growth, which suggests the existence of a turbulent alpha-effect operating in the convection zone, in agreement with former studies.
Second, this result is very sensitive to the boundary conditions. While a vertical field, or open boundary condition, allows the growing of the magnetic fields to values around 10\% of the equipartition value with kinetic energy of turbulent convection, in the case with perfect conductor (PC)  boundary the saturation of the magnetic field occurs earlier and with a smaller value \cite{Guerrero and de Gouveia Dal Pino2010}.

\begin{figure}[!htb]
\centering
\includegraphics[height=5cm]{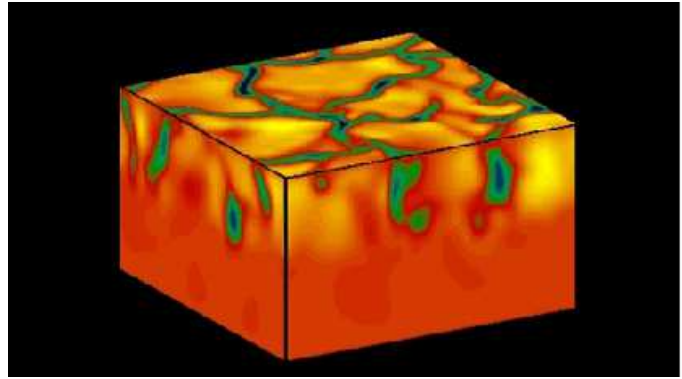}

  \caption{3-D magneto-convectin simulation of a cubic parcel of the convective (top) and radiative (bottom) layers of the Sun (from Guerrero \& de Gouveia Dal Pino 2010 \cite{Guerrero and de Gouveia Dal Pino2010}).}
\end{figure}

\begin{center}
\begin{figure}[!ht]
\includegraphics[height=6cm]{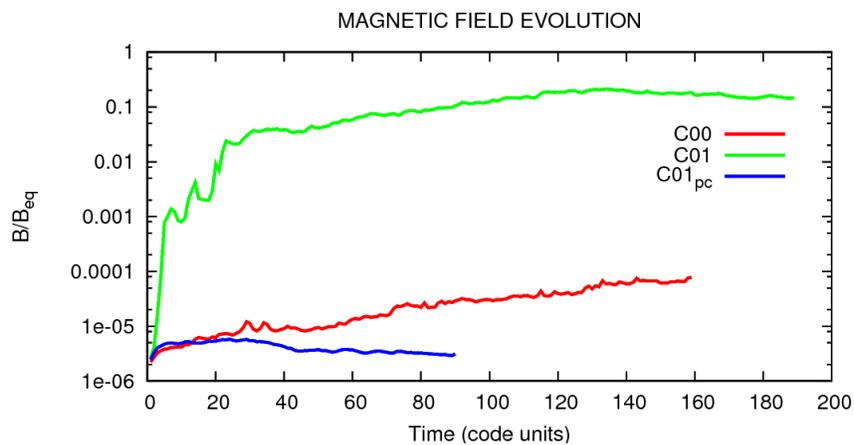}
  \caption{Magnetic field evolution for different simulated models. The growth of the magnetic field is sensitive to both the presence of rotation and the boundary condition:  with rotation and open boundary, it
  allows the growing of the magnetic fields to values around 10\% of the equipartition (green line),
  while a perfect conductor boundary with rotation leads to earlier saturation of the growth (blue line). An open boundary with no-rotation also leads to smaller growth (red line) (from Guerrero \& de Gouveia Dal Pino 2010 \cite{Guerrero and de Gouveia Dal Pino2010}).}
\end{figure}
\end{center}

\subsection{Magnetic Fields in other stars}

 Similar physical processes are expected to occur in other classes of stars, especially in the cool, convective ones. For recent reviews on the magnetism across  the HR diagram I refer to Berdyugina \cite{Berdyugina2009} and Jardine \& Donati \cite{Jardine and Donati2009}.

 In the last 5 to 10 years it has become apparent that the
change in internal structure with mass along the main sequence can have profound effects
on the types of magnetic fields that can be generated. At the high-mass end of the main
sequence, the presence of significant magnetic fields (of the order of 100 G) on some stars with radiative interiors
challenges current dynamo theories that rely on convective processes. One possibility is
that they generate fields in their convective cores, although this raises the question of
how to transport the
flux to the surface. They can also generate magnetic fields in the radiative zone, but a  non-solar
dynamo process would be required. Alternatively, the fields may be fossils, left
over from the early stages of the formation of the star (\cite{Jardine and Donati2009} and references therein).

Very low mass stars also have an internal structure that is very different from that
of the Sun in that convection may extend throughout their interiors. In the absence
of a tachocline, these stars cannot support a solar-like interface dynamo, yet they, like
the high mass stars, exhibit observable magnetic fields ($10-10^3$ G). The mechanism by which they
generate these fields has received great attention lately. While a
decade or so ago, it was believed that these stars could only generate small-scale magnetic
fields, more recent studies suggest that large
scale fields may be generated. These models differ, however, in their predictions for the
form of this field and the associated latitudinal differential rotation. They predict that
the fields should be either axisymmetric with pronounced differential rotation, or non-axisymmtric with minimal differential rotation (see \cite{Jardine and Donati2009} and references therein).


\section{Magnetic Fields in Jets and Accretion Disks}

Narrow conical (or cylindrical) supersonic jets  that channel mass, momentum, energy and magnetic flux from their parent sources to the outer ambient medium are observed in a wide variety of astrophysical objects,  from young stars (YSOs) and compact evolved objects (like galactic black holes or microquasars, and X-ray binary stars), to gamma-ray bursts (GRBs) and the nuclei of active radio galaxies (AGNs). Despite their different physical scales (in size, velocity, source mass, and amount of energy transported), they have strong morphological similarities that suggest a universal mechanism for their origin.
The currently most accepted model for jet production is based on the magneto-centrifugal acceleration out off a magnetized disk that surrounds the central source and accretes matter into it (\cite{Blandford and Payne1982}; see also \cite{de Gouveia Dal Pino2005} for a review on this model and jet phenomena).

There has been  observational and numerical evidences in favor of this model, but it does not explain  the quasi-periodic ejection phenomena
often observed in the different astrophysical jet classes, nor the origin of the large scale magnetic field which is required for this mechanism to operate at the accretion disk. The magnetic field could be either advected from the ambient medium by the infalling material or could be locally produced by dynamo action in the disk in a similar way to the one previously discussed for the sun and other stars (see \cite{de Gouveia Dal Pino et al2010} and references therein). In this regard, it has been  proposed that the observed X-ray and radio flares events accompanied by ejections of relativistic particles in relativistic jets (e.g., GRS 1915+105) could be produced by violent magnetic reconnection episodes just like on the solar surface. The process would take place when a large scale magnetic field is established either by turbulent dynamo in the accretion disk or by magnetic flux  advection from the outer disk regions \cite{de Gouveia Dal Pino and Lazarian2005, de Gouveia Dal Pino2006b, de Gouveia Dal Pino et al2010}.
In the case of microquasars and AGNs,  a diagram of the magnetic energy rate released by violent reconnection as a function of the black hole (BH)
mass spanning $10^9$ orders of magnitude shows that the magnetic reconnection power is more than sufficient to explain the observed
radio luminosities of the outbursts, from microquasars to low luminous AGNs. In the case of YSOs, a similar magnetic configuration can be reached
that could possibly produce observed x-ray flares in some sources and provide the heating at the jet launching base \cite{de Gouveia Dal Pino et al2010}.

\section{MAGNETIC FIELDS IN THE ISM AND STAR FORMATION REGIONS}
Observations of the polarized emission of the diffuse ISM of our Galaxy trace structures of pc and sub-pc sizes that carry valuable information about its turbulent nature. There is a straight correlation between turbulence, the average magnetic field and cosmic rays distribution, with their energy densities nearly in equipartition. The MHD turbulence distributes the energy from supernovae explosions, jets and winds from stars in the ISM of the Galaxy and the magnetic fields control the density and distribution of cosmic rays in the ISM and halo of the Galaxy.

While the magnetic fields in the diffuse ISM and in neutral hydrogen clouds (HI clouds) are predominantly turbulent, they tend to be much more regular within the denser molecular clouds where stars form. Polarization maps of the cores of these clouds are consistent with strong regular fields, from $\sim$ 100  G to few mG \cite{Crutcher2005}. The observations also indicate thermal to magnetic pressure ratios $P_{th}/P_B \sim   0.29-0.04$, and turbulent to magnetic pressure ratios $P_t/P_B  \sim  1.3-0.7$, in the diffuse ISM and the molecular clouds cores, respectively. These figures indicate a predominance of compressible turbulence on the dynamics of the diffuse ISM and of the magnetic fields on the molecular clouds, and suggest the following potential scenario for structure and star formation in the ISM: molecular clouds are possibly formed by the aggregation of HI clouds. Thus, compressible turbulence must dominate the aggregation process over both gravity and magnetic fields, sometimes forming self-gravitating clouds. Since the magnetic fields are strong enough even in the diffuse ISM medium, the aggregation must be primarily along magnetic flux tubes as magnetic pressure resists to compression in the direction  perpendicular to the field.
Residual effects of turbulence are seen in self-gravitating molecular clouds, but they appear to be magnetically supported, with the condition  $\rho g  \simeq  \nabla P_B$ approximately satisfied. When a cloud core becomes gravitationally unstable, magnetic  field flux diffusion by turbulent reconnection \cite{SantosLimaetal2010a} and later ambipolar diffusion of the neutral component of the gas through the magnetic field lines (e.g. \cite{Heitsch et al2004}) must drive its gravitational collapse and star formation in a fast (few free-fall times) timescale. The magnetic field will be also essential for removing the angular momentum excess from the protostellar cloud during this process, causing the so called magnetic braking that will conduct the protostar to its final collapse into a stable star (\cite{Crutcher2005, Li and Shu1996}).

\section{MAGNETIC FIELDS IN THE MILKY WAY AND OTHER GALAXIES}

As generally observed in other galaxies with spiral morphology, the large scale magnetic field in our galaxy, the Milky Way, seems to approximately follow the density spiral arms. However, Faraday rotation measurements (RM) from pulsars radio emission suggest the existence of multiple reversals along the galactic radius that have not been detected so far in any external galaxy \cite{Han2009}. To account for large-scale reversals, a bisymmetric magnetic spiral would be required (a configuration that is not predicted by a simple dynamo process). It is also possible that at least some of these field reversals are not of galactic extent, but due to local field distortions or magnetic loops of the anisotropic turbulent field component. In summary, the pattern of the large-scale regular magnetic field of the Milky Way is still a matter of debate (e.g., \cite{Beck2005}). A combination of cosmic-ray energy density measurements with radio synchrotron data yields a local strength of the total field of the order of 6  $\mu$G and 10  $\mu$G in the inner Galaxy, which are similar to the values in other galaxies (see below).

Large-scale spiral patterns of regular field are observed in grand-design, flocculent, and even in some irregular galaxies. In grand-design galaxies the regular fields are aligned parallel to the optical spiral arms, with the strongest regular fields (highest polarization) in inter-arm regions, sometimes forming magnetic spiral arms between the optical arms. This is an indication that the magnetic fields do not require the assistance of  the density spiral arms to develop.
Regular fields are strongest in inter-arm regions ($B \sim 15  \mu$G), while turbulent fields are strongest within the spiral arms ($B \simeq 20  \mu$G) due to the presence of intense star formation, stellar winds, jets, and SN shocks there. Processes related to star formation tangle the field lines, so that smaller polarization is observed in star-forming regions. The existence of large scale regular magnetic fields in the inter-arm region is suggestive of shear amplification by differential rotation just like in the solar dynamo.

Most of the field structures in the galaxies require a superposition of several dynamo modes to explain their origin and amplification (e.g., \cite{Vallee2004}. The turbulent dynamo \cite{Subramanian1998} considers a standard dynamo model acting in different scales of multiple cells and these micro-dynamos combine to build a large scale magnetic field. In some cases (e.g., M31), Faraday rotation of the polarized radiation reveals patterns which are signatures of large-scale regular, coherent fields in the galactic disks that could not be generated by direct compression or stretching of turbulent fields in gas flows. The turbulent dynamo mechanism would be able to generate and preserve these large scale coherent magnetic fields with the appropriate spiral shape (\cite{Beck2005} and references therein). Also, the presence of magnetic arms between gas arms in most of the spiral galaxies, provides an extra support for the global dynamo hypothesis; for if B were produced essentially by stellar winds, jets and SN explosions then one would expect much more intense fields within the arms where these phenomena are more frequent.

A survey of several dozens of spiral galaxies (see \cite{Beck2005, Beck2009} and references therein) has revealed average magnetic fields $<B> = 9 \mu$G. In starburst (SB) galaxies, i.e., galaxies with high rates of star formation, the average
field intensity is even larger, $B \simeq 30-50 \mu$G and in the nuclear regions of these galaxies (e.g., NGC1097), magnetic fields $B \simeq 100 \mu$G have been detected. This suggests a correlation between the magnetic fields and star formation in these galaxies. The magnetic fields may play an important role in the deflagration of SBs and, since these galaxies are believed to be the progenitors of spheroid galaxies, in the formation of spheroid galaxies \cite{Totani1999}.

The magnetic energy density in the inner disk of some galaxies has been found to be larger than the thermal energy density, comparable to that of turbulent gas motions (another result that is consistent with dynamo action), and dominant in the outer disk (e.g., NGC6946). These evidences led Battaner and Florido \cite{Battaner and Florido2000} to investigate the dynamics of these galaxies and they have found that magnetic field forces at the outside regions cannot be neglected in these cases and are a competitive candidate to explain the observed flatness of rotation curves without the necessity of invoking the presence of dark matter.

In some galaxies with high star formation rate it has been detected the presence of magnetic filaments and loops that are coupled with charged dust and rise above the galactic disk into the halo up to heights of  5 kpc or more (e.g., NGC891; Dettmar 2005). They resemble the magnetic structures observed at the solar surface and suggest the presence of similar magnetic activity. Supernova shock fronts can break through the galactic disk and inject hot gas into the galactic halo, perhaps driving turbulence and the magnetic fields into it  (e.g., \cite{Melioli and de Gouveia Dal Pino2004, Melioli et al2008, Melioli et al2009, Kim et al2006, Falceta et al2010}).
In several galaxies, the halo gas is observed to rotate slower than the gas in the disk (NGC5775). This velocity gradient could contribute to the excitation of a global dynamo in disk galaxies in a similar way to the differential rotation in the solar dynamo \cite{Dettmar2005}.

\section{MAGNETIC FIELDS IN CLUSTERS OF GALAXIES AND THE IGM}

Looking into the large scale structure of the Universe, we see that galaxies tend to aggregate into clusters. The Milky Way, for instance, belongs to the Virgo Cluster. A cluster has a typical diameter of  100 Mpc (3 million light-years) and can be regular or irregular.

Faraday rotation measurements of polarized Synchrotron radiation from embedded or background radio galaxies indicate magnetic fields in the Abel clusters $B \sim 2 \mu$G with coherent length scales $L \sim 10$ kpc. In the central regions of clusters with embedded radio galaxies the magnetic fields are even larger: $B \sim 5 \mu$G in irregular clusters, and  $B \sim 10-30 \mu$G in regular ones \cite{Govoni2006}, which are of the order of the magnetic field strengths of the galaxies. In some cluster cores, there seems to be some evidence that the magnetic pressure dominates the thermal pressure. Observations also reveal that the magnetic fields in the intra-cluster medium are predominantly turbulent with a power law spectrum which is approximately Kolmogorov at least at the small scale structures ( $\sim$ 1 kpc) (e.g., Hydra Cluster; \cite{Ensslin et al2005, Govoni2006}).

These magnetic fields may have been powered by galactic winds from starburst (SBs) galaxies, jets from radio galaxies, cluster mergers, or dynamo action (\cite{Ensslin et al2005, Vallee2004} and references therein). It has been proposed that a dynamo operating in the accretion disk around the massive black hole (BHs) in the nucleus of radio galaxies could produce magnetized jets and inject them into the intra-cluster medium. Numerical Simulations  \cite{Kato et al2004} of magnetic field generation from accretion disks have shown that this is possible. However, the dynamics of the process leading to the formation of massive BHs in the nuclei of the active galaxies is still unclear and a pre-existing magnetic field may be required to carry away the huge angular momentum of the accreting matter. Besides, radio sources can generate only $B \sim 10^{-7}$ G \cite{Kronberg2005}, so that further amplification would be required and could be provided by, e.g., shocks produced by galaxy interactions. Likewise, SB galaxies could also produce only $B \sim 5 \times 10^{-9}$ G that should be amplified by other mechanisms \cite{Kronberg et al1999}. Alternatively, Ensslin, Vogt \& Pfrommer \cite{Ensslin et al2005} have tried to explain the fields in core of clusters in terms of turbulent dynamo action, where instead of differential rotation the flow is in a turbulent medium. In this model, dynamo action occurs in all scales and micro-dynamos will compose to develop large scale magnetic fields. Magnetic diffusivity constrains the magnetic field scale to a fraction of the turbulent scale $L_B \propto  L_T (RM)^{-1/2}$, where $RM$ is the magnetic Reynolds number. As B grows, it reacts and tends to untangle thus decreasing $RM$ and increasing $L_B$ and generating a more organized B.

MHD turbulence fed by star formation and SN explosions have also been invoked  to explain the loops and filaments of gas extending to over 50 kpc around the central galaxy of the Perseus cluster \cite{Falceta et al2010}. 2.5 and three-dimensional MHD simulations have revealed that the turbulence injected by SNe could be responsible for the nearly isotropic distribution of filaments and loops that drag magnetic fields upward. Weak shell-like shock fronts propagating into the intracluster medium (ICM) with velocities of 100-500 km/s are found, also resembling the observations. As the turbulence is subsonic over most of the simulated volume, the turbulent kinetic energy is not efficiently converted into heat and additional heating is required to suppress the cooling flow at the core of the cluster. Simulations combining the MHD turbulence with the AGN outflow can reproduce the temperature radial profile observed around the central galaxy (See Figure 3). While the AGN mechanism is the main heating source, the SNe are crucial to $isotropize$ the energy and magnetic field distribution in the cluster core.

\begin{center}
\begin{figure}
\includegraphics[height=6cm]{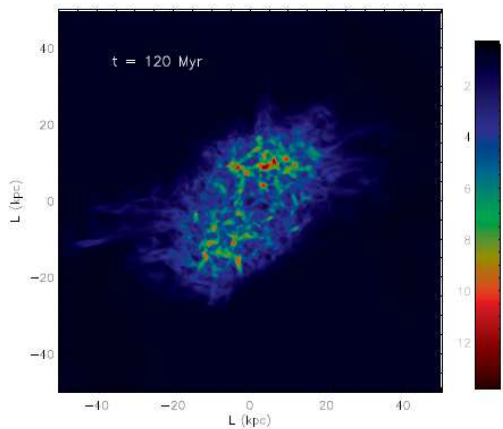}

  \caption{3D MHD simulation of the central core of Perseus cluster. This map shows the projected magnetic pressure ratio  distribution ($\int_{LOS}(B^2/B_0^2) dl/L$). During the formation of
the filaments, the magnetic field is dragged and compressed with the
gas. The local density within the filaments is $n\sim 0.01 - 0.04$
cm$^{-3}$, a factor of $\sim 10 - 100$ times larger than the ambient value. As a result,
the magnetic pressure within the overdense filaments and loops is typically larger than
the surroundings by a factor of $30 - 200$, i.e. the filaments present
absolute values of magnetic field intensities in the range $B_{\rm
fil} = 5 - 20 \mu$G (from Falceta-Gon\c{c}alves et al. 2010 \cite{Falceta et al2010}).}
\end{figure}
\end{center}

Magnetic fields probably pervade the entire Universe. The diffuse intergalactic medium (IGM) is very rarefied but Faraday rotation of polarized emission from distant quasars (up to z=2-6) indicate that $B_{IGM} \leq  10^{-9}$ G, assuming a coherent length of the order of the maximum one measured in clusters $L  \sim$ 1 Mpc (e.g., \cite{Grasso and Rubinstein2001}). Where and how do these magnetic fields originate? Are they primordial?

\section{MAGNETIC FIELDS IN THE EARLY UNIVERSE}

The history of the Universe started about 15 billion years ago with the Big-Bang singularity and since then it has expanded and cooled down, and suffered several phase transitions. When its temperature was about $10^9$ K and its size  100 pc, the primordial nucleosynthesis took place with the formation of the first H and He nuclei. When the Universe was about 0.5 Myr old, its temperature decreased to 10,000 K and the electrons and protons recombined to form neutral H atoms and the radiation decoupled from matter. This relic radiation from the Big-Bang has been cooling since then and presently permeates the whole cosmos with a microwave background radiation (CMBR) of temperature $T \simeq 3$ K. The recombination was followed by a dark age where possibly the first objects in the Universe were formed and these caused a reionization of it when it was about 0.1 Gyr old. This epoch was probably particularly favorable for creation of magnetic fields (see below). The reionization was followed by the formation of the galaxies and the larger structures in the Universe (at  $\sim$1 Gyr).

Both the CMBR and the Big-Bang nucleosynthesis (BBN) impose constraints on the strength of primordial fields (e.g., \cite{Shaposhnikov2005}; \cite{Grasso and Rubinstein2001}). In the case of the BBN, the presence of a strong magnetic field could change the expansion rate of the Universe and thus the abundance of the primordial elements. The observed cosmic helium abundance implies an upper limit for the magnetic field at the time of the BBN, $B < 10^{11}$ G which expanded to the present Universe (under the hypothesis of magnetic flux conservation and performing a proper statistical average of small flux elements; e.g. \cite{Grasso and Rubinstein2001}) would imply a present field permeating the IGM,  $B_o < 10^{-10}$ G (for coherent lengths of 1 Mpc). In the case of the CMBR, the existence of a strong magnetic field at the recombination era would have affected it through two effects: (i) breaking the spatial isotropy; and (ii) producing MHD alterations on the temperature and polarization fluctuations of the CMBR. These fluctuations have been imprinted on the CMBR by primordial density fluctuations (probably produced in the Universe right after the B-B, during the inflation phase, and later on allowed the formation of the large structures in the Universe). The presence of a strong magnetic pressure could, for example, reduce the acoustic peaks of these fluctuations by opposing the infall of photons and baryons into their gravitational well. The lack of these effects on the CMBR implies an upper limit on the background magnetic field at the present Universe $B_o < 10^{-8} -10^{-9}$ G. Thus, these upper limits derived from the CMBR and BBN constrain the present cosmic magnetic field to values that are consistent with the inferred values for the IGM today (10$^{-9}$ G.).

Also, it is possible to demonstrate that the diffusion time scale of these primordial fields is much larger than the age of the Universe and the corresponding diffusion length of these magnetic fields is $< 10^9$ cm, so that any magnetic fields produced in the early Universe with present length scales larger than 10$^9$ cm would have survived and probably left no significant imprints on the BBN or CMBR!

Is the inferred IGM magnetic field above primordial? As remarked in the previous sections, the origin of the galactic and cluster magnetic fields is probably due to the amplification of seed fields via galactic dynamo or more recent processes. But where and when were these first fields produced?

The existing models for primordial magnetic fields are still rather speculative. There are those that propose more recent astrophysical origins for the magnetic fields and their seeds. Among these, one can mention the Biermann battery in intergalactic shocks, the Harrison effect \cite{Harrison1970}, stellar magnetic fields, supernova explosions, galactic outflows into the intergalactic medium, and jets from active galaxies; and there are those models called cosmological ones, that invoke sometimes peculiar processes in the very early Universe (prior to the recombination era) to create seed magnetic fields. I will mention below only some examples of these models (and recommend, e.g., \cite{Grasso and Rubinstein2001, Shaposhnikov2005, de Gouveia Dal Pino2006a, Rees2006};  for those interested in  more complete reviews of the subject).

Among more recent, post-recombination candidates, the Biermann battery appears as a promising mechanism for seed field generation. It is well known that a third term should appear in the magnetic induction equation (eq. 1) whenever the flow has pressure and density gradients which are non parallel . During the re-ionization era, this mechanism may have created pre-galactic seed fields $B_o \simeq 10^{-21}$ G  that were then exponentially amplified by dynamo action within the proto-galactic clouds that originated the galaxies later on. Another possibility is the Harrison mechanism, originally proposed by Harrison (1970) to operate in proto-galactic structures, it could also have operated in  pre-recombination plasmas where vortices would be present. The essential idea of this mechanism is that in a primordial flow with vortical structures, protons will rotate (through collisions) with the flow, but the electrons will be Thomson scattered by the photons of the CMBR.  This will cause charge separation and generate electric currents and thus magnetic fields. It can be shown that the operation of this mechanism within vortices in the pre-recombination plasma would be able to create seeds with $10^{-18}$ G with correlation lengths of 1 Mpc in the present IGM. These could be amplified by dynamo action within the galaxies and provide the observed galactic magnetic fields ( few  $\mu$G). A recent interesting suggestion, which is alternative to the dynamo mechanism, is that seed fields could be amplified by supernova-driven turbulence. Kim et al. (2006) argue that the observation of magnetic fields of  $\mu$G levels already in very distant  galaxies which were formed very long ago (at high redshifts, $z \simeq$  2) shorten the time available for dynamo action.  In order to amplify a seed field by the alpha-omega dynamo mechanism, which requires few Gyr growth time, the strength of the seed field should be stronger than 10$^{-11}$ G. On the other hand, the growth time of magnetic fields amplified by SN-driven turbulence is only $\sim$10 Myr, for 100 pc scales. This suggests a more efficient mechanism for kpc-scale magnetic field amplification, which should, however, still be confirmed by numerical experiments with large computational domains.

Recently, a new paradigm for the investigation of cosmic magnetic fields in the IGM has appeared. There is growing evidence that the plasma in the ICM and IGM has low density ($n \sim  10^{-3}$ cm$^{-3}$) and is weakly collisional, i.e., the ion Larmour frequency around the magnetic field is much larger than the ion-ion collision frequency (\cite{Shekochihin2005}). In this case, the thermal pressure becomes anisotropic with respect to the magnetic field orientation and the evolution of the turbulent gas is more correctly described by a kinetic MHD approach (KMHD).
Assuming the pre-existence of seed fields generated, e.g., by the Weibel instability, the Harrison mechanism, or the Biermann battery in proto-galactic clouds, or even more recent mechanisms, like the injection of small-scale fields into the IGM and ICM produced in star-forming galaxies, active radio galaxies, and galaxy mergers, several groups are now  investigating how these fields can be amplified and made coherent by turbulence evolution (e.g., \cite{Shekochihin2005}; \cite{SantosLimaetal2010b}). These studies must shade new light on the structure and evolution of cosmic fields in the coming years.

As an example, Figures 4 and 5 show preliminary tests made with our  kinetic MHD  (KMHD) Godunov based code \cite{SantosLimaetal2010b}. Performing 3D numerical simulations of non-helical, forced turbulence considering  the presence of an initial seed magnetic field, Figures 4 and 5 compare models in the  KMHD regime with those of a collision MHD plasma, we have found that the KMHD produces a more folded field distribution. The instabilities (e.g., fire-hose) that develop due to the anisotropic pressure in this case accumulate energy at the smallest scales, changing the inclination of the power spectrum of the amplified magnetic field and producing small-scale fluctuations in the field that make it more wrinkled than in the case of a collisional KMHD plasma (see Figures 4 and 5).  Further simulations including parallel heating conduction, cooling and more realistic thresholds for the instabilities must still be performed for a complete study of the dynamo action in KMHD. We are currently working on these new tests \cite{SantosLimaetal2010b}.

\begin{center}
\begin{figure}[!htb]
  \includegraphics[height=7cm]{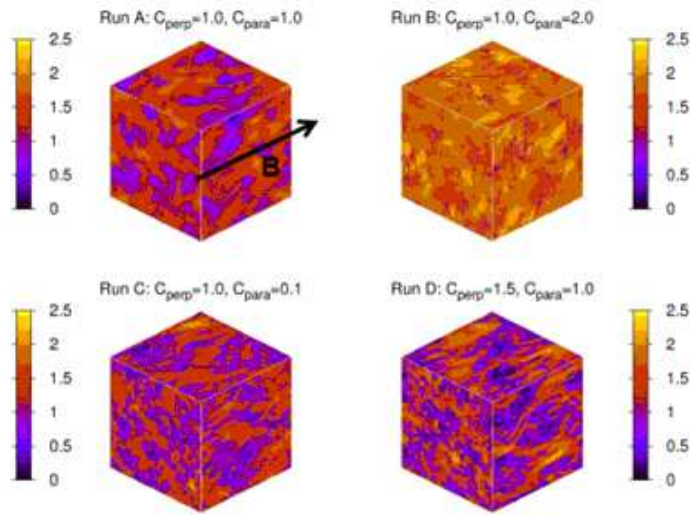}
  \caption{Magnetic field distribution for turbulent intergalactic medium (IGM) with distinct initial values of the perpendicular and parallel sound speeds to the direction of the magnetic field ($C_{perp}$, $C_{par}$). The top left panel is for a pure MHD case, while the others are   KMHD cases. We note that the larger the degree of unisotropy the more wrinkled the system is. (from Santos-Lima et al. 2010b \cite{SantosLimaetal2010b}).}
\end{figure}
\end{center}

\begin{figure}[!htb]
\centering
  \includegraphics[height=6cm]{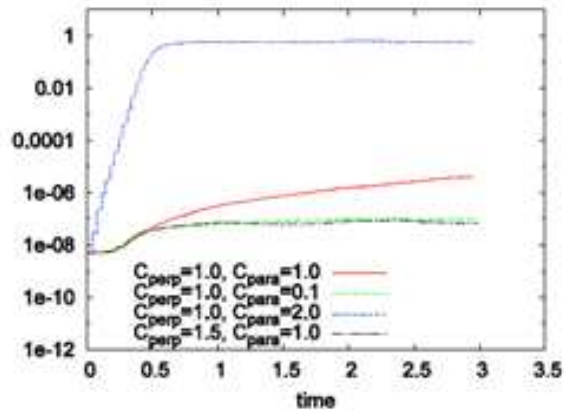}
  \caption{Magnetic energy evolution for the different systems of the models of Figure 4. Only the system that is in the regime which is unstable to the fire-hose instability shows an efficient amplification of the seed field.  (from Santos-Lima et al. 2010b \cite{SantosLimaetal2010b}).}
\end{figure}

\section{FUTURE NEEDS AND PERSPECTIVES}

Advances are needed not only in the theory and magneto-hydrodynamical multidimensional numerical studies, but perhaps even more urgent is the requirement for polarization observations at higher angular resolution in order to map the full wealth of magnetic structures in galaxies, the ISM and the IGM, that will in turn provide inputs for dynamo and MHD models. In our Galaxy, for instance, we need polarization maps of selected regions at high frequencies and a much larger sample of rotation measure data from pulsars (\cite{Beck2005, Gaensler2009}).

There is presently a project for the construction of a next-generation radio telescope array that will have a total effective collecting area of 1 km$^2$ (with sensitivity at the frequencies 100 MHz to 25 GHz), the so called Square Kilometer Array (SKA). It will be composed of about 150 stations of 100 m diameter - accounting for half the SKA area -  that will be distributed across continental distances ($\sim$3000 km), and the remaining area (called core) will be concentrated within a central region of 5 km diameter. It will be built under an international collaboration involving several European countries. Cosmic magnetism is one of the key science projects of the SKA. It will be able to detect more than 10,000 pulsars in our Galaxy which will allow to map the large scale spiral structure of its magnetic field. Besides, the SKA will be able to map nearby galaxies with at least 10 times better angular resolution compared to present-day radio telescopes, or 10 times more distant galaxies with similar spatial resolution as today. Magnetic field structures will illuminate the dynamical interplay of cosmic forces. The SKA sensitivity will allow to detect synchrotron emission from the most distant galaxies in the earliest stage of evolution and to search for the earliest magnetic fields and their origin (e.g., \cite{Gaensler2009}).

SKA is planned to be built only around 2020. Meanwhile, a new generation of instruments is already in operation or under construction that will allow the investigation of cosmic magnetism in all scales, from compact and stellar sources to the more diffuse ISM and IGM. Among these I should mention GALFACTS (operation since 2008 in Arecibo), LOFAR (consisting of arrays with sensitivity in two small frequency bands: 20-80 MHz and 110-240 MHz that will be operating in Netherlands and Germany in 2010); ATA (operating in California since 2008);
SKAMP (planned to operate in Canberra in 2010); MWA (to become available in Western Australia in 2010); EVLA (in New Mexico in 2010); MeerKAT (in South Africa in 2012); and ASKAP (in Western Australia in 2012). At millimetric and submillimetric  frequency ranges there are the ALMA (Atacama Large Millimetric Array) and the LLAMA (Large Latin American Millimetric Array) arrays, which are planned to be operating in Chile in 2015, and in Argentina/Brazil in 2011, respectively.

\begin{theacknowledgments}
  The author is indebted to all her students and collaborators who made possible the production of this review.

\end{theacknowledgments}

\end{document}